\documentclass[aps,pra,preprint]{revtex4}

\usepackage{times}

\usepackage{amsmath}
\usepackage{amsfonts}
\renewcommand{\baselinestretch}{1}

\begin{document}

\title{Entanglement in Non-Hermitian Quantum Theory}
\author{Arun K.\ Pati \footnote{
Invited plenary talk in the International Conference 
(Homi Bhabha Centenary Conference) on Non-Hermitian Hamiltonians in 
Quantum Physics (PHHQP VIII) held at BARC, Mumbai during Jan 13-16, 2009.
}}
\affiliation{Institute of Physics, Bhubaneswar-751005, Orissa, India}



\begin{abstract}

Entanglement is one of the key feature of quantum world that has no
classical counterpart. This arises due to the linear superposition
principle and the tensor product structure of the Hilbert space when we deal
with multiparticle systems. 
In this paper, we will introduce the notion of
entanglement for quantum systems that are governed by non-Hermitian yet 
$PT$-symmetric 
Hamiltonians. We will show that maximally entangled states in usual quantum 
theory behave like non-maximally entangled states in $PT$-symmetric 
quantum theory. Furthermore, we will show how to create entanglement 
between two $PT$Qubits using non-Hermitian 
Hamiltonians and discuss the entangling capability of such interaction
Hamiltonians that are non-Hermitian in nature. 

\end{abstract}

\maketitle

\def\ra{\rangle}
\def\la{\langle}
\def\ver{\arrowvert}



\section{ Introduction}
Entanglement is one of the weirdest feature of quantum mechanics. 
In quantum world entanglement arises naturally when we have more 
than two particles at our 
disposal. There is no classical analog of quantum entanglement and that makes 
it more fascinating than anything else in physics. Though, there is a burst of
activity in understanding the nature of entanglement, the concept by itself
is not new. It was 
introduced by Schr{\"o}dinger way back in 1935 and he has realized that 
{\it ``entanglement  is the characteristic trait 
of quantum mechanics, the one that enforces its entire departure from classical
lines of thought}'' \cite{sch}.
In the emerging field of quantum information theory
entanglement plays a major role. This is also a very useful resource in the 
sense that using entanglement one can do many things in the quantum world
which are usually impossible in ordinary classical world. Some of these tasks 
are quantum computing \cite{qc}, quantum teleportation \cite{qt}, 
quantum cryptography \cite{qcr}, remote state preparation \cite{akp},
quantum communication \cite{quc}, and so on. 
The fundamental carrier of information in quantum 
world is a quantum bit or qubit. A qubit is any two-state quantum 
mechanical system that can exist simultaneously in both $0$ and $1$.
It differs from a classical bit in many ways. Some important differences 
are that we cannot copy a qubit \cite{wz,dd} nor can we delete a qubit 
from two identical copies \cite{pb}.

In standard quantum mechanics the 
observables are represented by Hermitian operators and the evolution of 
a closed system is governed by unitary evolution.
In recent years there is a considerable interest in quantum systems governed by
non-Hermitian Hamiltonians \cite{bend1,bend2,bend3,most,most1,most2}.
It was discovered that there are class of non-Hermitian Hamiltonians, 
yet they posses real eigenvalues provided they respect 
$P$T symmetry and the symmetry is unbroken. 
In $PT$-symmetric quantum mechanics the usual condition 
of Hermiticity of
operators is replaced by the condition of $CPT$ invariance, 
where $C$ stands 
for charge conjugation, $P$ for parity and $T$ for time reversal \cite{bend1}. 
In standard quantum theory $CPT$ symmetry and Hermiticity 
conditions are the same. The $CPT$ invariance condition is a natural 
extension of 
Hermiticity condition that allows reality of observables and unitary dynamics.
Using the operator $C$, Bender et al \cite{bend2} have introduced an inner 
product structure 
associated with $CPT$ which can have positive definite norms for quantum states.

In this paper we would like to introduce the notion of 
entanglement for quantum systems described by non-Hermitian Hamiltonians.
Usually, with non-Hermitian Hamiltonians one may think that there will be 
dissipation in the system and one may not be able to create entanglement. 
But, we will show how can we create entanglement with interaction 
Hamiltonians that are non-Hermitian in nature.
Towards the end, we will address 
what is the entangling capability of non-Hermitian interaction Hamiltonians.
Before doing so, first we will give basic definitions of entanglement in
standard quantum theory. Then we will introduce the notion of $PT$-symmetric 
quantum bit (PTQubit) and the notion of quantum 
entanglement in this theory. Because of the $CPT$ inner product, orthogonal 
quantum states in ordinary quantum theory become non-orthogonal quantum 
states in non-Hermitian quantum theory. This has several consequences 
which will be explored in detail.
Also, we will show that if we take an Einstein-Podolsky-Rosen (EPR) entangled 
state (which is known to be a maximally entangled state) in ordinary 
theory that becomes a non-maximally entangled state in non-Hermitian quantum 
theory. Towards the end 
some implications and open questions will be discussed. 
We hope that the entanglement 
in $PT$-symmetric quantum theory may provide new ways of 
processing information in the quantum world.

\section{Entanglement in usual quantum theory}
Let us consider a composite system that consists of two or more subsystems.
The Hilbert space of a composite system is the tensor product of the 
individual Hilbert spaces. In the case of bipartite quantum system we have 
the joint Hilbert space ${\cal H} = {\cal H}_1 \otimes {\cal H}_2$.
If the state of a composite system cannot be written as 
$|\Psi \rangle_{12} = |\psi \rangle_1 \otimes |\phi \rangle_2$, then 
it is an entangled state. Suppose
 $\{ |\psi_n \rangle \} \in  {\cal H}_1^N$ and 
$\{ |\phi_m \rangle \} \in {\cal H}_2^M$ are the basis in the respective 
Hilbert spaces, then 
$\{ |\psi_n \rangle_1 \otimes |\phi_m \rangle_2 \} \in 
{\cal H}_1^N \otimes {\cal H}_2^M$ is a basis in the joint Hilbert space. 
A general pure bipartite state can be expressed as 
\begin{eqnarray}
|\Psi \rangle_{12} = \sum_{nm=1}^{NM} C_{nm} 
|\psi_n \rangle_1 \otimes |\phi_m \rangle_2.
\end{eqnarray}
The above state cannot be written in product form for general amplitudes, 
hence it is an entangled state. Thus, a generic pure bipartite state is actually
an entangled  state.
There is a beautiful theorem called the Schmidt decomposition 
theorem which tells that any 
pure bipartite entangled state can be written as 
\begin{eqnarray}
|\Psi \rangle_{12} = \sum_{i=1}^{ min (N, M)} \sqrt{p_i} 
|a_i \rangle_1 \otimes |b_i \rangle_2,
\end{eqnarray}
where $p_i \ge o $ are the Schmidt coefficients and
$|a_i \rangle, |b_i \rangle$ are the Schmidt vectors, and $\sum_i p_1=1$.
It can be seen that if we have more than one non-zero Schmidt coefficients 
in the bipartite state then it is an entangled. The Schmidt coefficients are 
invariant under local unitary transformations.

Now, if we want to define the state of the individual systems, then they 
are given by partial traces, i.e., 
\begin{eqnarray}
\rho_1 &=& {\rm tr}_2 (|\Psi\rangle_{12}{}_{12}\langle \Psi|) 
= \sum_i p_i |a_i \rangle \langle a_i | \nonumber \\ 
{\rm and}~~~\rho_2 &=& {\rm tr}_1 (|\Psi\rangle_{12}{}_{12}\langle \Psi|)
= \sum_i p_i |b_i \rangle \langle b_i|.
\end{eqnarray}
Note that $\rho_1$ and $\rho_2$ are no longer pure, i.e., 
$\rho_i^2 \not= \rho_i (i=1,2)$. This is another indication that the original
state of the composite system is an entangled state. If it is not, then after 
performing partial trace the reduced density matrices will be still pure. 
The existence of the Schmidt decomposition for bipartite states guarantees that 
the reduced density matrices have equal spectrum, though the eigenvectors can be
different. It may be stated that if we have an entangled state of three or 
more particles then there does not exist a Schmidt decomposition. The 
necessary and sufficient conditions for the existence of 
Schmidt decomposition  was found in Ref. \cite{arun}.
If $A$ is a linear Hermitian operator acting on ${\cal H}_1$ and 
if $B$ is a linear Hermitian operator acting on ${\cal H}_2$, then
the expectation values of these local observables are given by 
\begin{eqnarray}
{}_{12}\langle \Psi|A \otimes I |\Psi\rangle_{12} &=& {\rm tr}_1 (\rho_1 A),
\nonumber\\
{\rm and}~~~ {}_{12}\langle \Psi|I \otimes B |\Psi\rangle_{12} &=& 
{\rm tr}_2 (\rho_2 B).
\end{eqnarray}
This suggests that the expectation values of the local observables are 
completely  determined by local (reduced) density matrices.

For any pure bipartite state one can quantify how much entanglement is there in 
a given state. The entropy of any one of the reduced density matrix 
is a very good measure of 
entanglement for any bipartite state $|\Psi\rangle$ \cite{pr}.
It is given by  

\begin{eqnarray}
E(\Psi ) = 
- {\rm tr}_1(\rho_1 \log \rho_1) =  - {\rm tr}_2(\rho_2 \log \rho_2) 
= - \sum_i p_i \log p_i.
\end{eqnarray}
This measure of entanglement satisfies the following properties:\\
\noindent
(i) $E(\Psi) = 0$ iff $|\Psi \ra$ is separable.\\
(ii) $E(\Psi)$ is invariant under local unitary transformations, i.e., 
$E(\Psi) = E(U_1 \otimes V_2 \Psi)$.\\
(iii) $E(\Psi)$ cannot increase under local operation and classical 
communications (LOCC).\\
(iv) The entanglement content of $n$ copies of $|\Psi\ra$ is additive, i.e., 
$E(\Psi^{\otimes n} ) =  n E(\Psi)$.\\

The above ideas can be illustrated with two qubits and two-qudits 
(qudit is a $d$-dimensional Hilbert space system) entangled states.
One famous entangled state which has been extensively used in quantum 
information theory is the Einstein-Podolsky-Rosen (EPR) \cite{epr} state
$|\Psi^-\rangle$ which is given by
 \begin{eqnarray}
|\Psi^-\rangle = \frac{1}{\sqrt 2}(|0 \rangle|1 \rangle - 
|1 \rangle|0 \rangle).
\end{eqnarray}
This has one unit of entanglement or one (entangled bit) ebit (because 
$\rho_1 = \rho_2 = I/2 $). This is also a 
maximally entangled state for two-qubits. In fact, any state which is locally 
equivalent to $|\Psi^-\rangle$ will have one unit of entanglement. 
Similarly, in a higher dimensional
Hilbert space ($d\times d$) a maximally entangled state for two-qudits 
 can be written as 
\begin{eqnarray}
|\Phi\rangle = 
\frac{1}{\sqrt d} \sum_{i=0}^{d-1} |i \rangle \otimes |i \rangle
\end{eqnarray}
which has has $E(\Phi)= \log d$ ebits. Here also any other state such as
$(U_1 \otimes V_2)|\Phi\rangle$  will have $\log d$ ebits of entanglement, where
$U_1$ and $V_2$ are local unitary operators acting on ${\cal H}_1$ and 
${\cal H}_2$, respectively.

In information theory (both classical and quantum) there is a famous slogan due 
to Landauer: {\it ``Information is physical''}.
In the same spirit, I would like to say that {\it Entanglement is Physical}. 
This is justified
for the following reasons: Entanglement can be created, stored, and consumed 
using physical systems and physical operations.
Entanglement is independent of any particular representation. For example, one
ebit can be stored in two photons, two electrons or two atoms.
As said before, entanglement is a resource. One can do informational work like 
quantum computing, 
quantum teleportation, remote state preparation, 
quantum cryptography and many more.

Since I am not going to review all the details of entanglement here, let me 
mention some recent trends in entanglement theory.
For last several years, characterization and quantification of entanglement 
of multiparticle system is a vigorous area of research \cite{plenio}.
Understanding of how well one can generate entanglement is another direction 
scientists are exploring. Also, there is an upsurge of interest in 
understanding the dynamics of entanglement. In this context many authors have 
investigated entanglement rate and entangling capabilities of
non-local Hamiltonians \cite{dur}, entangling power of quantum 
evolutions \cite{zana}, various entangling operations \cite{cirac}, 
and simulation of one Hamiltonian 
by another using only local operations \cite{bennett} and so on.

\section{Non-Hermitian Quantum Theory}

In this section we will give the basic formalism that is necessary to develop
the notion of entanglement in non-Hermitian quantum theory.
Recently, there has been a great deal of interest in studying 
$PT$-symmetric quantum theory \cite{bend1,bend2,bend3,most,most1,most2}. 
In earlier formulation of 
$PT$-symmetric quantum theory, it turned out that $PT$-symmetric quantum theory
introduced 
states which have negative norms. This had no clear interpretation. 
This was cured by introducing another operator $C$ called conjugation 
operator \cite{bend1,bend2}. This operator commutes with the Hamiltonian 
and the operator $PT$. Also $C^2=I $, which implies that it has 
eigenvalues $\pm 1$ 
.

Bender {\it et al} \cite{bend1,bend2} have shown that non-Hermitian 
Hamiltonians can have real eigenvalues if it possess 
$PT$-symmetry, i.e., $[H, PT]= 0$  and the symmetry is unbroken 
(if all of the eigenfunctions of $H$  are simultaneous eigenfunction of 
the operator $PT$). Hamiltonians having unbroken $PT$ symmetry can 
define a unitary quantum theory.
Unitarity can be shown by the fact that such Hamiltonians possess a new 
symmetry called conjugation $C$ with $[C, H]=0$ and $[C, PT]=0$.

Quantum theory that deals with non-Hermitian Hamiltonians and 
respects $CPT$ symmetry may be called non-Hermitian 
quantum theory. One can formalise this by stating the following postulates:\\

\noindent
(i) A quantum system is a three-tuple $({\cal H}, H,  
\langle . | . \rangle_{CPT} )$, where ${\cal H}$ is a physical Hilbert 
space with the $CPT$ inner product $\langle . | . \rangle_{CPT}$ having a 
positive norm, and $H$ is the non-Hermitian Hamiltonian. \\

\noindent
(ii) The state of a system is a vector $|\psi \rangle$ in ${\cal H}$.
 For any two vectors the $CPT$ inner product is defined as
 $\langle \psi | \phi \rangle_{CPT} = 
\int~ dx [ CPT \psi(x)] \phi(x)$.\\

\noindent
(iii) The time evolution of state vector is unitary with respect to $CPT$
inner product.\\

\noindent
(iv) An observable can be a linear operator $O$, provided it is Hermitian 
with respect to the $CPT$ inner product, i.e., 
$\la .|O~. \ra_{\rm CPT} = \la O~ .| .\ra_{\rm CPT} $.\\

\noindent
(v) If we measure an observable $O$, then the eigenvalues are the possible 
outcomes.\\

\noindent
(vi) If measurement gives an eigenvalue $O_n$, the states makes a transition 
to the eigenstate $|\psi_n\ra$ and the probability of obtaining the 
eigenvalues $O_n$ (say) in a state 
$|\psi\rangle$ is given by 
\begin{eqnarray}
p_n =\frac{ |\langle \psi | \psi_n \rangle_{CPT} |^2 }
{||\psi||_{CPT} ||\psi_n||_{CPT} },
\end{eqnarray}
where $||\psi ||_{\small CPT} = \sqrt{\langle \psi | 
\psi \rangle_{\small CPT} }$.\\

\noindent
(vii) If we have two quantum systems $({\cal H}_1, H_1,  
\langle . | . \rangle_{CPT} )$ and 
$({\cal H}_2, H_2,  
\langle . | . \rangle_{CPT} )$, then the state of the combined system will 
live in a tensor product Hilbert space ${\cal H}_1 \otimes {\cal H}_2$.\\

Some remarks are in the order. In our effort to introduce entanglement we are
using $CPT$ inner product and the above postulates. However, one can also use 
the pseudo-Hermiticity approach \cite{most,most1} 
and do similar thing. Incidentally, the 
physical observable was defined as the one that is invariant under $CPT $
operation \cite{bend2}. It was shown to be inconsistent with the 
dynamics of the theory 
\cite{most2}. Then, it was modified and suggested that 
an observable should satisfy $O^{T} = (CPT) O (CPT)$, where $O^{T}$ is the 
transposition of $O$.
This guarantees that the expectation value of $O$ in any state is real. 
However, this definition restricts that Hamiltonian be not only $PT$-symmetric 
but also symmetric \cite{most3}.

\section{$PT$-Symmetric Quantum Bit}

In standard quantum mechanics, we say that any two-state system is a 
quantum bit or a qubit. For example, an arbitrary state of a spin-half 
particle like $|\Psi \ra = \alpha |\uparrow \ra + \beta |\downarrow \ra$ 
can represent a qubit.  Here, $|\uparrow \ra$ and $|\downarrow \ra $ are 
the eigenstates of the Pauli matrix $\sigma_z$.
Similarly, if we have a two-level atom, then an 
arbitrary superposition of the ground state and the first excited state will be 
a qubit. In fact, any arbitrary superposition of two orthogonal
states can represent a qubit.
In the same vein, in $PT$-symmetric quantum mechanics if
we store information in any two-state system, then we call it as a 
$PT$-symmetric quantum bit or in short {\em PTQubit}.
In general $PT$Qubit is different from a qubit.

In non-Hermitian quantum theory a general two-state system 
will be described by a $2\times 2$ Hamiltonian which respects $CPT$ symmetry.
Following the Ref.\cite{bend1}, this Hamiltonian is given by

\begin{eqnarray}
H = \left( \begin{array}{rr} r~e^{i\theta} & s \\  t~~ &   r~e^{-i\theta} 
\end{array} \right),
\end{eqnarray}
with $r, s, t,$ and $\theta$ all are real numbers. This Hamiltonian is 
non-Hermitain yet it has real eigenvalues whenever we have $st > r^2 \sin^2 
\theta$. Also, this Hamiltonian is invariant under $CPT$. Two distinct 
eigenstates of this Hamiltonian are given by 
\begin{eqnarray}
|\psi_+\ra = \frac{1}{\sqrt 2 \cos \alpha} 
\left( \begin{array}{r} e^{i\alpha/2} \\  e^{-i\alpha/2} \end{array} \right)~~ 
{\rm and}~~
|\psi_-\ra = \frac{1}{\sqrt 2 \cos \alpha} 
\left( \begin{array}{r} e^{-i\alpha/2} \\  -e^{i\alpha/2}  
\end{array} \right),
\end{eqnarray}
where $\alpha$ is defined through $\sin \alpha = \frac{r}{\sqrt{st}} 
\sin \theta$. With respect to the $CPT$ inner product (which gives a positive 
definite inner product) we have 
$\la \psi_{\pm} | \psi_{\pm} \ra_{\rm CPT} = 1$ and 
$\la \psi_{\pm} | \psi_{\mp} \ra_{\rm CPT} = 0$. The $CPT$ inner product for 
any two states of $PT$Qubit is given by
\begin{eqnarray}
\la \psi|\phi \ra = [(CPT) |\psi \ra]. \phi, 
\end{eqnarray}
where $\la \psi|$ is the $CPT$ conjugate of $|\psi\ra$.
In the $2$-dimensional Hilbert space, the operator $C$ is given by 
\begin{eqnarray}
C = \frac{1}{\sqrt 2 \cos \alpha} \left 
( \begin{array}{rr} i\sin \alpha & 1 \\  1  &  -i\sin \alpha  
\end{array} \right).
\end{eqnarray}
The operator $P$ is unitary and is given by
\begin{eqnarray}
P =  \left ( \begin{array}{rr} 0 & 1 \\  1  &  0 \end{array} \right).
\end{eqnarray}
The operator $T$ is anti-unitary and its effect is to transform 
$x \rightarrow x, p \rightarrow -p$ and
$i \rightarrow -i$.  

Since the eigenstates $|\psi_{\pm}\rangle$ of the non-Hermitian Hamiltonian 
$H$ span the 
two-dimensional Hilbert space, one can encode
one bit of information in these orthogonal states. 
An arbitrary state can be 
represented as superposition of these orthogonal states 
\begin{eqnarray}
|\Psi \ra = \alpha |\psi_+ \ra + \beta |\psi_-\ra = 
\alpha |0_{\rm CPT} \ra + \beta |1_{\rm CPT}\ra.
\end{eqnarray}
Thus, any arbitrary 
superposition of two orthogonal states of $PT$ invariant Hamiltonian will be
called $PT$-quantum bit or {\it PTQubit}. In fact, any linear superposition 
of two orthogonal states of an observable $O$ in $PT$-symmetric quantum 
theory can represent a {\it PTQubit}.

\section{Entanglement in Non-Hermitian Theory}

Entanglement is one of the most important feature of quantum world \cite{epr}. 
As noted earlier, when we 
have more than one qubit then the state of the composite system may be 
found in an entangled state that has no classical analog. 
Now, in $PT$-symmetric quantum theory we will have similar feature whenever we 
have more than one $PT$qubit. In this section, we introduce these basic notions.

Suppose we have two quantum systems with non-Hermitian 
Hamiltonians $H_1$ and $H_2$, where
\begin{eqnarray}
H_1 = \left( \begin{array}{rr} r e^{i\theta} & s \\  s~~ &   r e^{-i\theta} 
\end{array} \right)~ {\rm and}~~
H_2 = \left( \begin{array}{rr} r'e^{i\theta'} & s' \\  s'~~ &   
r'e^{-i\theta'} \end{array} \right).
\end{eqnarray}
Let  $\{ |\psi_{\pm}\rangle \} \in {\cal H}_1$ and 
$\{ |\psi'_{\pm}\rangle \} \in {\cal H}_2$ are the eigenfunctions of the 
Hamiltonians $H_1$ and $H_2$, respectively.
Now, the state of the combined system will live in ${\cal H}_1 \otimes 
{\cal H}_2$ which is spanned by $\{ |\psi_+\rangle \otimes 
|\psi_+'\rangle, |\psi_+\rangle \otimes |\psi_-'\rangle, |\psi_- \rangle 
\otimes |\psi_+'\rangle, |\psi_-\rangle \otimes |\psi_-'\rangle \}$. 
If the combined state cannot be written as 
$|\Psi\rangle = |\psi\rangle \otimes |\phi\rangle = 
|\psi\rangle |\phi\rangle $, then it is entangled.
A general state of two $PT$qubit can be expanded using the joint basis 
in ${\cal H}_1 \otimes {\cal H}_2$ as
\begin{eqnarray}
|\Psi\rangle = a |\psi_+\rangle \otimes |\psi_+'\rangle 
 + b |\psi_+\rangle \otimes |\psi_-'\rangle + c |\psi_-\rangle 
\otimes |\psi_+'\rangle
+ d |\psi_-\rangle \otimes |\psi_-'\rangle.
\end{eqnarray}
For general values of the complex amplitudes $a, b, c$ and $d$ 
this is an entangled state.
However, if $\frac{a}{b} = \frac{c}{d}= k$, then $|\Psi\rangle$ 
is not entangled. Now, we can quantify the entanglement content 
in $|\Psi\rangle$. It is given by the entropy of the reduced state 
of any one of the subsystem, i.e, 

\begin{eqnarray}
E(\Psi) = -\lambda_+ \log \lambda_+ 
- \lambda_- \log \lambda_-,
\end{eqnarray}
where 
$\lambda_{\pm} = \frac{1}{2}(1 \pm \sqrt{X})$ and 
$X = 1 - 4[(|a|^2 + |b|^2)(|c|^2 + |d|^2) - |(ac^* + b d^*)|^2 ]$.
For $\frac{a}{b} = \frac{c}{d}= k$, $E(\Psi) = 0$, as expected.

Now, the $CPT$ inner product on the Hilbert spaces ${\cal H}_1$ and 
${\cal H}_1$ can be used to define the inner product on 
${\cal H}_1 \otimes {\cal H}_1$.
For any two arbitrary vectors $|\Psi\rangle, |\Phi \rangle 
\in {\cal H}_1 \otimes {\cal H}_2$, we define the inner product between them 
as   

\begin{eqnarray}
\langle \Psi | \Phi \rangle_{CPT} = 
[(CPT) \otimes (CPT ) |\Psi\rangle]. |\Phi \rangle.
\end{eqnarray}
Using this inner product we can calculate relevant physical quantities for 
the composite system under consideration.

One can generalize the notion of entanglement for more than two $PT$qubits. 
If we have $n$-$PT$Qubits with individual Hamiltonians as 
$H_i (i=1,2, \cdots n)$ 
with respective eigenbasis $\{ |\psi_{\pm i} \ra \}$, then the joint Hilbert 
spaces will be ${\cal H}_1 \otimes {\cal H}_2 \cdots \otimes {\cal H}_n$. 
If a joint
state cannot be written as 
 $|\psi \ra_1 \otimes |\phi \ra_2 \cdots \otimes |\chi \ra_n$, 
then it will be an 
entangled state. A general $n$-$PT$Qubit state can be written as
\begin{eqnarray}
|\Psi \ra = \sum_{k=0}^{2^n -1} \alpha_k |X_k \ra,
\end{eqnarray}
where $|X_k \ra$ is a $n$-bit string of all possible combinations of 
$|\psi_{\pm}\ra$. Such a states will be generically an entangled state.
However, in this paper we are not going to dwell upon 
multi $PT$Qubit systems.

In general, if we have two subsystems with non-Hermitian Hamiltonians in
higher dimension (${\cal H}^d \otimes {\cal H}^d$), then we can also 
introduce the notion of entanglement. A general state of two $PT$-symmetric 
quantum systems 
can be written as (note that for non-Hermitian quantum systems also 
we can write a Schmidt decomposition theorem)
\begin{eqnarray}
|\Psi \rangle = \sum_{i} \sqrt{\lambda_i} 
|\psi_i \rangle \otimes |\phi_i \rangle. 
\end{eqnarray}
Now the reduced states of the $PT$-symmetric particles $1$ and $2$ will be 
different if we calculate the partial traces in usual quantum theory and 
in non-Hermitian quantum theory.
Because the inner products in ordinary and $PT$-symmetric quantum theory 
are different, the partial traces will also be different. For example, 
the reduced density matrix for particle $1$ calculated in non-Hermitain 
quantum theory will be 
\begin{eqnarray}
\rho_1 & = &\sum_{ij} \sqrt{\lambda_i \lambda_j }
|\psi_i \rangle \langle \psi_j | {\rm tr}_2(|\phi_i \rangle \langle \phi_j |) =
\sum_{ij} \sqrt{\lambda_i \lambda_j }
|\psi_i \rangle \langle \psi_j | [(CPT) |\phi_j \rangle] . |\phi_i \rangle  
\nonumber \\
& = & \sum_{i} \lambda_i |\psi_i \rangle \langle \psi_i |.
\end{eqnarray}
But if we calculate the reduced density matrix of particle $1$ in usual 
quantum theory, then we will have 
\begin{eqnarray}
\rho_1 = \sum_{ij} \sqrt{\lambda_i \lambda_j }
|\psi_i \rangle \langle \psi_j | {\rm tr}_2(|\phi_i \rangle \langle \phi_j |) =
\sum_{ij} \sqrt{\lambda_i \lambda_j }
|\psi_i \rangle \langle \psi_j | \la \phi_j |\phi_i \rangle.
\end{eqnarray}
This is no more in diagonal form because 
$\la \phi_j |\phi_i \ra \not= \delta_{ij}$ in the usual sense. 
Similarly, one can check that 
the reduced density matrix of the particle $2$ will be different in two 
theories. The density for particle $2$ in non-Hermitain theory will be 
\begin{eqnarray}
\rho_2 & = &\sum_{ij} \sqrt{\lambda_i \lambda_j }
|\phi_i \rangle \langle \phi_j | {\rm tr}_1(|\psi_i \rangle \langle \psi_j |) =
\sum_{ij} \sqrt{\lambda_i \lambda_j }
|\phi_i \rangle \langle \phi_j |~[(CPT) |\psi_j \rangle] . |\psi_i \rangle  
\nonumber \\
& = & \sum_{i} \lambda_i |\phi_i \rangle \langle \phi_i |.
\end{eqnarray}
But in the usual 
quantum theory, we will have 
\begin{eqnarray}
\rho_2 = \sum_{ij} \sqrt{\lambda_i \lambda_j }
|\phi_i \rangle \langle \phi_j | {\rm tr}_1(|\psi_i \rangle \langle \psi_j |) =
\sum_{ij} \sqrt{\lambda_i \lambda_j }
|\phi_i \rangle \langle \phi_j | \la \psi_j |\psi_i \rangle.
\end{eqnarray}
As a consequence, the entanglement content of a bipartite state depends on the 
inner product being used to calculate the partial traces. In other words, 
$E(\Psi)= S(\rho_i)~ (i=1,2) $ in usual quantum theory is not equal to 
$E(\Psi)= S(\rho_i)~ (i=1,2) $ in the 
non-Hermitian quantum theory.

To illustrate the above idea, 
we can define a singlet state for two $PT$qubits as 
\begin{eqnarray}
|\Psi_{\rm CPT}^- \rangle = \frac{1}{\sqrt 2}(|\psi_+ \rangle
| \psi_- \rangle - | \psi_-\rangle | \psi_+ \rangle.
\end{eqnarray}
In $PT$-symmetric quantum theory, the entanglement content of 
$|\Psi_{\rm CPT}^- \rangle$ is given by 
$E(\Psi_{\rm CPT}^-) =1 $. Note that
this is not the usual spin singlet $|\Psi^- \rangle$.
This is because the entanglement content of  $|\Psi^- \rangle$ 
in non-Hermitian quantum theory will be different.

This is one interesting aspect here. The singlet state in ordinary quantum 
theory has entanglement equal to one whereas in $PT$-symmetric quantum
theory it will be less than one. Similarly, a singlet state in $PT$-symmetric
quantum theory will have 
entanglement equal to one whereas in ordinary theory it will be less
than one. This is because of different nature of the inner products in ordinary 
and non-Hermitian quantum theory.
To see this clearly, let us consider the entangled state of spin-singlet in 
ordinary quantum theory. If we want to know the entanglement content in 
$PT$-symmetric quantum 
theory then we have to calculate the von Neumann entropy of the reduced 
density matrix in $PT$-symmetric theory. The reduced 
density matrix for particle $1$ in non-Hermitian quantum theory is given by 

\begin{eqnarray}
\rho_1 = {\rm tr}_2(|\Psi^- \ra \la \Psi^- |) = \frac{1}{2} 
[|0\ra \la 0| \la 1|1\ra_{\rm CPT} - |0\ra \la 1| \la 0|1\ra_{\rm CPT} 
- |1 \ra \la 0 | \la 1|0\ra_{\rm CPT}  + |1 \ra \la 1| \la 0|0\ra_{\rm CPT} ], 
\nonumber \\
\end{eqnarray}
where the $CPT$ inner products are given by 
$\langle 0 |0 \rangle_{\rm CPT} = \langle 1|1\rangle_{\rm CPT}  
= \frac{1}{\cos \alpha}, 
\langle 0 |1 \rangle_{\rm CPT} = i \tan \alpha$ and  
$\langle 1 |0 \rangle_{\rm CPT} 
= -i \tan \alpha$. Using this the reduced density matrix for particle $1$
is given by
\begin{eqnarray}
\rho_1 = {\rm tr}_2(|\Psi^- \rangle \langle \Psi^- |) = 
\frac{1}{2 \cos^2 \alpha} 
\left 
( \begin{array}{rr} 1+\sin^2\alpha~  & -2i \sin \alpha \\  
2i \sin \alpha~  &  1 +\sin^2\alpha
\end{array} \right).
\end{eqnarray}
Note that $\rho_1$ is not normalized. We can define a normalized density 
matrix ${\tilde \rho}_1 = \rho_1/{\rm Tr } \rho_1 $, so that 
\begin{eqnarray}
{\tilde \rho}_1 = \frac{1}{2} 
\left 
( \begin{array}{rr} 1~~~  & -2i \sin \alpha \\  
2i \sin \alpha  &  1 
\end{array} \right).
\end{eqnarray}

The eigenvalues of the density matrix ${\tilde \rho}_1$ are given by
$\lambda_{\pm} = \frac{1}{2}(1 \pm 2 \sin \alpha)$
Now, the entanglement content of usual singlet in $PT$-symmetric quantum theory 
is given by 
\begin{eqnarray}
E(\Psi^-) = -\lambda_1 \log \lambda_1 - \lambda_2 \log \lambda_2 = 
- \frac{1}{2}(1 + 2\sin \alpha) \log \frac{1}{2}(1 + 2 \sin \alpha) \nonumber \\
- \frac{1}{2}(1 - 2 \sin \alpha) \log \frac{1}{2}(1 - 2 \sin \alpha) \not= 1.
\end{eqnarray}

This shows that if an entangled state in ordinary theory has one 
unit of entanglement, in non-Hermitian quantum theory it will have less than 
one unit of entanglement. 
This is the effect of non-Hermiticity on the quantum entanglement. 
In the Hermitian limit ($\alpha =0$), $E(\Psi^-) = 1$.
Similarly, one can check that the maximally entangled state 
$|\Psi_{\rm CPT} \ra = \frac{1}{2}(|\psi_+ \ra |\psi_-\ra - |\psi_- \ra 
|\psi_+\ra) $ of non-Hermitian 
quantum theory will have less than one unit of 
entanglement in ordinary quantum theory. One implication of such an effect is
that if Alice and Bob share an EPR entangled state generated by $PT$-symmetric 
quantum world, then they cannot use that for quantum teleportation in 
ordinary world. Because, 
perfect quantum teleportation requires one ebit of entanglement.

\section{Generation of Entanglement with non-Hermitian Hamiltonian}

We know that entanglement can be created between two systems via 
some interaction. In standard quantum theory, interactions are described by 
Hermitian Hamiltonians. One might think that with non-Hermitian Hamiltonians, 
one may tend to destroy entanglement. However, here we show how to create 
entanglement with such  non-Hermitian Hamiltonians.

A general Hamiltonian for two-particles in $PT$-symmetric quantum theory  
is given by $H = H_1 \otimes I_2 + I_1 \otimes H_2 + H_{12}$, where 
$H_1, H_2$ and  $H_{12}$ could be non-Hermitian but respect $PT$ symmetry. 
Total Hamiltonian must satisfy $[H, PT\otimes PT] =0$.
If $|\Psi(0) \rangle = |\psi(0) \rangle \otimes 
|\phi(0) \rangle$ evolves to $|\Psi(t) \rangle$ under the action of this 
non-local Hamiltonian, then the state at a later time could be entangled, i.e.,

\begin{eqnarray}
|\Psi(t) \rangle =  
e^{-i Ht} |\psi(0) \rangle \otimes |\phi(0) \rangle \not= 
|\psi(t) \rangle \otimes |\phi(t) \rangle
\end{eqnarray}

An important question is what is the best way to exploit the interaction 
to produce entanglement? First we will give a simple 
non-local Hamiltonian that is capable of creating entanglement. 
Consider an interacting Hamiltonian given by
\begin{eqnarray}
H = \left( \begin{array}{rr} r e^{i\theta} & s \\  s~~ &   r e^{-i\theta} 
\end{array} \right) \otimes 
\left( \begin{array}{rr} r'e^{i\theta'} & s' \\  s'~~  &   r'e^{-i\theta'} 
\end{array} \right)
\end{eqnarray}
which satisfies $[H, PT\otimes PT]=0$.
Using the Pauli matrices we can write $H$ as 
\begin{eqnarray}
H = [r \cos \theta I + \frac{\omega}{2} \sigma. {\bf n}] \otimes
[r' \cos \theta' I + \frac{\omega'}{2} \sigma. {\bf n'}]
\end{eqnarray}
where ${\bf n} = \frac{2}{\omega}(s, 0, ir \sin \theta)$, 
$ \omega = 2 s \cos \alpha$, 
similarly for ${\bf n}'$ and $\omega'$. This interaction Hamiltonian 
consists of local terms and non-local terms. To see this we write it 
explicitly as
\begin{eqnarray}
H = rr' \cos \theta \cos \theta' (I\otimes I) +  
r \cos \theta \frac{\omega'}{2} (I \otimes \sigma. {\bf n'}) 
 + r' \cos \theta'  \frac{\omega}{2} (\sigma. {\bf n} \otimes I)
+ \frac{\omega \omega'}{4} (\sigma. {\bf n} \otimes \sigma. {\bf n'}).
\end{eqnarray}
In the above expression, first, second and third terms are local terms.
We know that the local terms cannot create entanglement, so 
they can be transformed away. Only term
which is capable of creating entanglement is $\frac{\omega \omega'}{4} 
(\sigma. {\bf n} \otimes \sigma. {\bf n'})$.
Therefore, the 
entangling evolution operator is given by 
\begin{eqnarray}
U(t) &=& \exp[ -i \frac{\omega \omega' t}{4} 
(\sigma. {\bf n} \otimes \sigma. {\bf n'}) ] \nonumber\\
&=& \cos \frac{\omega \omega' t}{4} I - i \sin \frac{\omega \omega' t}{4}
(\sigma. {\bf n} \otimes \sigma. {\bf n'}).
\end{eqnarray}

If the initial state of two $PT$Qubit $|\Psi(0)\rangle = 
| 0 \rangle \otimes  |0 \rangle$, then 
at a later time $t$ the state is given by  
\begin{eqnarray}
|\Psi(t)\rangle &=& 
e^{ -i \frac{\omega \omega' t}{4} 
(\sigma. {\bf n} \otimes \sigma. {\bf n'}) } | 0 \rangle \otimes  |0 \rangle
\nonumber\\
&=& \alpha(t) | 0 \rangle |0 \rangle + \beta(t) | 0 \rangle |1 \rangle 
+ \gamma(t) | 1 \rangle   |0 \rangle + \delta(t) | 1 \rangle |1 \rangle.
\end{eqnarray}
where $\alpha(t)= \cos(\frac{\omega \omega' t}{4}) + 
i \sin(\frac{\omega \omega' t}{4}) \frac{4}{\omega \omega'} r r' \sin \theta 
\sin \theta'$, 
$\beta(t)=
\frac{4}{\omega \omega'} \sin(\frac{\omega \omega' t}{4})  s' r  \sin \theta$, 
$\gamma(t) = 
\frac{4}{\omega \omega'} \sin(\frac{\omega \omega' t}{4})  s r'  \sin \theta'$,
and $\delta(t) =
-i\frac{4 s s'}{\omega \omega'} \sin(\frac{\omega \omega' t}{4})$.
It is clear that for the above values of the amplitudes
 $|\Psi(t)\rangle$ is indeed an entangled state. Note that
$|\Psi(t)\rangle$ is not normalized as the initial state that we have chosen
is also not normalized (under $CPT$ inner product).

\section{Entangling Capability of non-Hermitian Hamiltonians}

Given an interaction Hamiltonian, what is the most efficient way of 
entangling particles? For Hermitian interaction Hamiltonians it is known 
that \cite{dur} 
(i) it is better to start with initial entangled state, 
(ii) the best initial entanglement is independent of the physical process, 
(iii) one can improve the capability if we allow fast local operations, and 
(iv) in some cases, the capability improves by using ancillas.
Now the question is whether similar facts hold for non-Hermitian 
Hamiltonians that respects $PT$ symmetry? In this section, we will define the
entanglement rate for non-Hermitian Hamiltonians. But we do not yet know
if all these holds for non-Hermitian case. It is plausible that the above 
facts may still hold.

Let an initial state $|\Psi(0)\rangle$ evolves to $|\Psi(t)\rangle$ 
via an interaction Hamiltonian $H$ which is non-Hermitian.
Now, $|\Psi(t)\rangle$ can be entangled and the ability to create entanglement 
depends
on the nature of interaction and on the initial state.
To quantify the entanglement production, define the entanglement rate 
$\Gamma(t) = \frac{dE(t)}{dt}$, where $E(t)= E(\Psi)$ is 
entanglement measure for the state $|\Psi(t)\rangle$.
For example, the entanglement measure can be the von Neumann entropy 
of the reduced density matrix.

Let the state of two $PT$Qubits at time $t$ is 
\begin{eqnarray}
|\Psi(t)\rangle = \sqrt \lambda_1(t) |a_1(t)\rangle |b_1(t)\rangle + 
\sqrt \lambda_2(t) 
|a_2(t)\rangle |b_2(t)\rangle
\end{eqnarray}
with $\langle a_1(t)|a_2(t)\rangle_{CPT} = 
\langle b_1(t)|b_2(t)\rangle_{CPT} =0$ and 
$\lambda_1 + \lambda_2 =1$. 
The amount of entanglement at time $t$ is given by the entropy of the reduced
density matrix (with  $\lambda = \lambda_1$)
 \begin{eqnarray}
E(\Psi(t)) 
= - \lambda(t) \log \lambda(t)
- (1- \lambda(t)) \log (1 - \lambda(t)).
\end{eqnarray}

The entanglement rate is given by 
\begin{eqnarray}
\Gamma(t) = \frac{dE(\Psi)}{d \lambda} \frac{ d \lambda}{dt}.
\end{eqnarray}
Using the Schr{\"o}dinger equation we have
\begin{eqnarray}
\frac{d  \lambda}{dt} = 2\sqrt{\lambda (1- \lambda)} {\rm Im} 
\langle a_1(t) | \langle b_1(t)|H|a_2(t) \rangle |b_2(t) \rangle_{CPT}
\end{eqnarray}
Therefore, the entanglement rate is
\begin{eqnarray}
\Gamma(t) = f(\lambda) |h(H, a_1, b_1)|,
\end{eqnarray}
where $f(\lambda) = 2\sqrt{ \lambda (1- \lambda)} \frac{dE}{d \lambda}$ 
and $h(H, a_1, b_1) = 
\langle a_1(t) | \langle b_1(t)|H|a_2(t) \rangle |b_2(t) \rangle_{CPT} $.
Let $h_{\rm max}$ is maximum value of $|h(H, a_1, b_1)|$. Then 
$h_{\rm max} = {\rm max}_{||a_1||, ||b_1|| =1 }
|\langle a_1(t) | \langle b_1(t)|H|a_2(t) \rangle |b_2(t) \rangle_{CPT}|$.
As in the Hermitian case, if we solve for $\frac{d\lambda}{dt}$, we have
$\lambda(t) = \sin^2(h_{\rm max} t + \phi_0), {\rm with}~  
\lambda_0 = \sin^2(\phi_0)$.
The evolution of entanglement is characterized by $h_{\rm max}$  which 
depends on the interaction Hamiltonian.
Thus, for a given $H$, $h_{\rm max}$ measures the capability of creating 
entanglement. 
The entanglement rate satisfies 
\begin{eqnarray}
\Gamma(t) \le \log [(1- \lambda)/\lambda]~  h_{\rm max},
\end{eqnarray}
showing that the bound is proportional to the entangling capability.

In future, we will investigate the entanglement rate for 
two entangled $PT$-symmetric quantum systems in higher dimension and see if all 
known results for Hermitian case also hold for non-Hermitian case.

\section{Conclusions}
In this paper we have 
introduced the notion of entanglement for quantum systems described by 
non-Hermitian Hamiltonians.
We have introduced the notion of
$PT$Qubit in the non-Hermitian quantum theory. 
Qubit states which are orthogonal in ordinary quantum theory 
become non-orthogonal in $PT$-symmetric quantum theory
 and vice verse. More interestingly, the entanglement property 
of quantum states also
change if we go from one theory to another. We have shown that 
a maximally entangled state that has von Neumann entropy 
equal to unit in the ordinary theory will have less entropy in $PT$-symmetric 
quantum theory and vice verse. One implication is that if 
there is a source that emits maximally entangled state in the sense of 
ordinary theory and two observers are now in non-Hermitian quantum world
then they cannot use the entangled state for quantum teleportation. This is 
because in their world the entanglement is not equal to unity. 
We have illustrated how to create entanglement between two $PT$Qubits using 
non-Hermitian Hamiltonians. We have discussed the entangling capability of 
interaction Hamiltonians that are non-Hermitian in nature. In future, we 
would like to apply these ideas in the context of entangled brachistocrone 
problem in $PT$-symmetric quantum theory. We hope that the fascinating field
of entanglement will take a new turn in the non-Hermitian quantum world. 
In particular, it will be interesting to see if $PT$-symmetric entanglement 
can offer something new for quantum information processing and in sharpening
our understanding of quantum channels.

Before ending, I would like to make the following remark. Early formulation 
of $PT$-symmetric quantum theory aimed to offer a genuine extension of 
usual quantum theory. Later, mathematical unitary equivalence has been 
shown between 
pseudo-Hermitian quantum theory and the usual quantum theory for single quantum
systems \cite{most}. However, entangled quantum systems may offer new 
insights into 
the nature of this equivalence. Because, equivalence property of 
entangled states are 
different under joint unitary and under local unitary transformations, I 
conjecture that under local unitary transformations (or more generally under 
LOCC paradigm) equivalence between pseudo-Hermitian and the usual quantum 
theory may not exists. One hopes to 
discover something new in such situations.\\

\renewcommand{\baselinestretch}{1}

{\it Note Added:} After completion of this work, A. Mostafazadeh informed me 
in Mumbai during the International Conference on 
Non-Hermitian Hamiltonians in Quantum Physics (Jan 13-16, 2009)   
about Ref. \cite{sola}, where compound systems 
have been described using pseudo-Hermitian quantum theory.\\

\end{document}